\documentclass[manuscript]{aastex}

\def\kms{$\rm km~s^{-1}$ }

\def\C3{$\rm C~III$}
\def\OO5{$\rm O~V$}
\def\N3{$\rm N~III$ }
\def\O6{$\rm O~VI$}
\def\fe18{$\rm [Fe~XVIII]$}
\def\Si12{$\rm Si~XII$}
\def\Al11{$\rm Al~XI$}
\def\si8{$\rm [Si~VIII]$}
\def\Fe10{$\rm [Fe~X]$}
\def\Ni14{$\rm [Ni~XIV]$}
\def\Ca14{$\rm [Ca~XIV]$}
\def\SSi3{$\rm Si~III$ }

\def\kms{$\rm km~s^{-1}$}

\begin{document}

\title{Carbon, Helium and Proton Kinetic Temperatures in a Cygnus Loop Shock Wave}
\shorttitle{Ion Temperatures in a Cygnus Loop Shock Wave}

\author{John C. Raymond,\altaffilmark{1}} 
\author{Richard J. Edgar,\altaffilmark{1}}
\author{Parviz Ghavamian\altaffilmark{2}}
	\and
\author{William P. Blair\altaffilmark{3} }

\altaffiltext{1}{Harvard-Smithsonian Center for Astrophysics, 60 Garden St., 
Cambridge, MA  02138, USA; jraymond@cfa.harvard.edu}
\altaffiltext{2}{Dept. of Physics, Astronomy \& Geosciences, Towson University, Towson, MD  21252}
\altaffiltext{3}{Department of Physics and Astronomy, Johns Hopkins University, 3400 N. Charles St., Baltimore, MD 21218, USA}

\begin{abstract}
Observations of SN1006 have shown that ions and electrons in the plasma behind fast supernova 
remnant shock waves are far from equilibrium, with the electron temperature much lower than the
proton temperature and ion temperatures approximately proportional to ion mass.  In the
$\sim$360 \kms\/ shock waves of the Cygnus Loop, on the other hand, electron and ion temperatures
are roughly equal, and there is evidence that the oxygen kinetic temperature is not far
from the proton temperature.  In this paper we report observations of the He II $\lambda$1640 line
and the C IV $\lambda$1550 doublet in a 360 \kms\/ shock in the Cygnus Loop.  While
the best fit kinetic temperatures are somewhat
higher than the proton temperature, the temperatures of He and C are consistent with the proton
temperature and the upper limits are 0.5 and 0.3 times the mass-proportional temperatures, implying
efficient thermal equilibration in this collisionless shock.  The equilibration of 
helium and hydrogen affects the conversion between proton temperatures determined from H$\alpha$
line profiles and shock speeds, and that the efficient equilibration found here reduces the shock
speed estimates and the distance estimate to the Cygnus Loop of Medina et al. (2014) to about 800 pc.


\end{abstract}


\section{Introduction}

It is often assumed that astrophysical shock waves produce Maxwellian velocity
distributions in the downstream plasma and that all particle species have the
same temperature.  However, in the modest number of cases where
temperatures of different species can be measured, this is usually
not the case. Shock waves in settings such as the solar wind, 
the interstellar medium, and galaxy clusters are collisionless, in that 
the shock transition occurs on a scale comparable to the proton gyroradius,
which is orders of magnitude smaller than the collisonal mean free path.
In this case, one might expect that a collisionless shock thermalizes a fraction of the 
kinetic energy of each incoming particle, leading to mass-proportional 
temperatures.  The actual situation is more complex. 

Typical shock waves in the solar wind have modest
Mach numbers, and they generally produce electron temperatures $T_e$ around 0.2 times the 
proton temperatures $T_p$, while heavier ions are preferentially heated,
$T_i > T_p$ \citep{ghavamian13, korreck07}.  In the higher Mach number shocks
in supernova remnants (SNRs), $T_e / T_p$ declines from about 1 to less than 0.1 as the
shock speed increases from about 350 \kms\/ to 2000 \kms\/ \citep{ghavamian01, ghavamian02,
hughes00, ghavamian07, vanadelsberg, medina}.  Ion temperatures in SNR shocks have been more
difficult to measure, but $T_i \simeq m_i T_p / m_p$ for helium, carbon and oxygen
in a 2000-3000 \kms\/ shock in SN1006 \citep{raymond95, laming96, korreck04,broersen13}, while the oxygen 
temperature is less than 1.7 times the proton temperature in a 350 \kms\/ shock 
in the Cygnus Loop \citep{raymond03}.    However, coulomb collisions affect the oxygen temperature
in the Cygnus Loop.  In the $\simeq$ 900 \kms\/ shocks in the LMC
supernova remnant DEM L71, the kinetic temperature of O derived from the O VI line widths
is about 6.5 times the proton temperature, though this is somewhat uncertain due to
interstellar absorption and the contribution of bulk motions to the line width \citep{ghavamian07b}. 
The shocks in galaxy clusters are as fast as the
shocks in young SNRs, but the pre-shock temperatures are so high that the Mach number
is only around 3.  Some seem to show rapid electron-ion temperature equilibration, while
others do not \citep{russell12}.

Direct measurements of velocities in solar wind shocks generally show
non-Maxwellian distributions \citep{thomsen90}.  Particle distributions in SNR shocks are
non-Maxwellian in the sense that they have power-law tails of relativistic
particles, but little is known about the velocity distribution of the vast majority
of particles in the core of the velocity distribution.  There is some evidence for a non-thermal
distribution at one position in Tycho's SNR, but it is ambiguous \citep{raymond10}.
Simulations with particle-in-cell codes
tend to show Maxwellian cores with power-law tails \citep{caprioli}.

In this paper we measure the kinetic temperatures of helium and carbon from
profiles of the UV lines of He II and C IV obtained with the COS spectrograph
on the Hubble Space Telescope.  We observed a 360 \kms\/ shock in the Cygnus Loop 
for which electron-ion temperature ratio is close to 1 \citep{medina}.  In order
to interpret the observations we needed the line profile of the COS instrument
with the G160M grating for an extended source, so we observed a position in the 
planetary nebula NGC 6853.  We also compute the effects of optical depth and Coulomb 
collisions on the line profiles.   We find that the helium and carbon kinetic 
temperatures, like the electron temperature, are close to the proton temperature.

\section{Observations and Data Reduction}

The target position in the Cygnus Loop coincides with the single orbit
COS spectrum labeled Pos 1 in \cite{raymond13} and the H$\alpha$
observation labeled COS1 in \cite{medina}.  Nine more
orbits were obtained with COS using the G160M grating set
at 1577 \AA ; 4 orbits on 2014 October 3 and 5 orbits on 2014 November 11.
They were combined with the earlier exposure for a total observation
time of 29,100 seconds.  The circular entrance aperture is 2.5$^\prime$$^\prime$
in diameter.

\begin{figure}
\epsscale{0.96}
\plotone{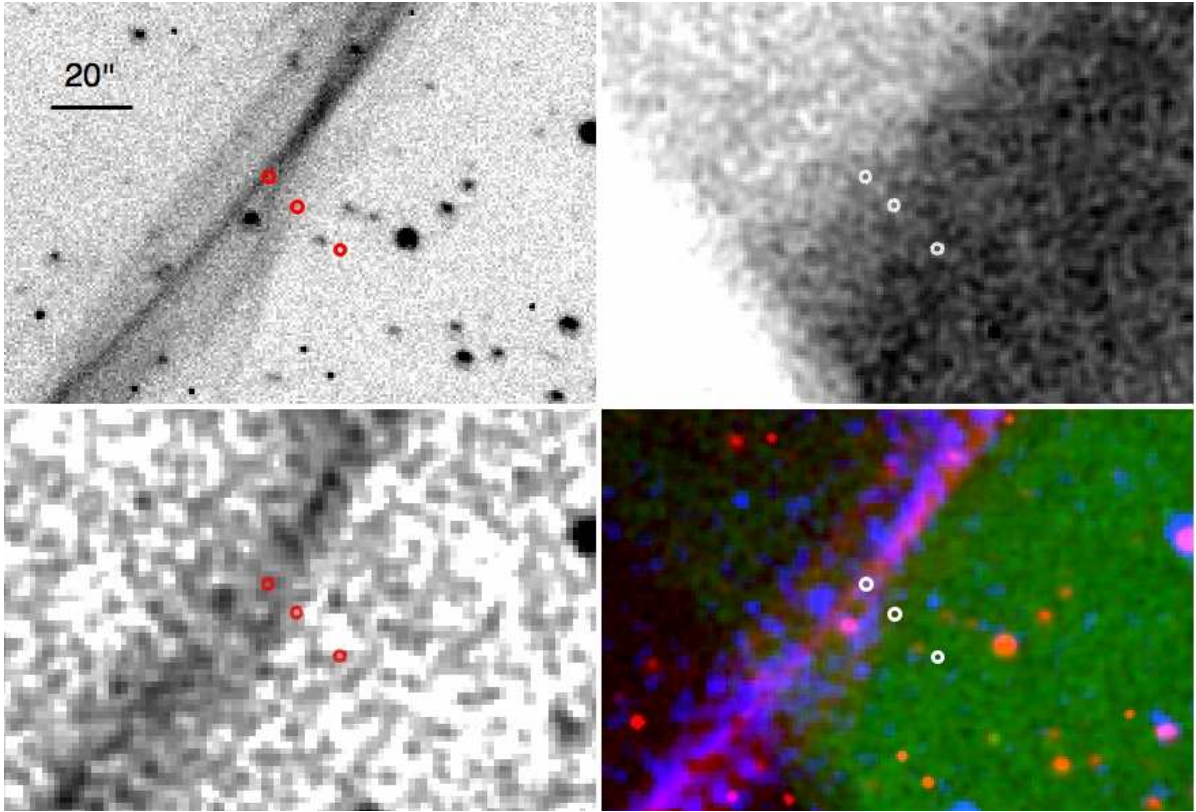}
\caption{ COS aperture positions of \cite{raymond13} overlaid on H$\alpha$,
{\it Chandra} X-ray and Galex NUV images.  The lower right hand
panel shows the apertures overlaid on a 3 color superposition
of H$\alpha$ (red), NUV (blue) and X-rays (green).
The position observed here is the one at the upper left on the H$\alpha$ filament.
}
\label{aperturepos}
\end{figure}

The position is shown in Figure~\ref{aperturepos}, which also shows
the two positions behind the shock reported in \cite{raymond13}.  It
was chosen to be slightly behind the brightest H$\alpha$ emission
based on the ionization times required to reach He II and C IV
in the shocked gas.  The coordinates are RA(2000) = 20$^h$ 54$^m$ 43.611$^s$,
Dec(2000) = +32$^\circ$ 16$^\prime$ 3.53$^\prime$$^\prime$.  For comparison, 
we include and X-ray image from {\it Chandra} and and NUV image from {\it Galex}.
The X-rays show that the shock coincides with the jump in electron temperature.
The structure in the {\it Galex} image is very similar to that seen in H$\alpha$.
This is probably because the {\it Galex} NUV passband from about 1900 \AA\/ to 2750 \AA\/ 
is dominated by the 2-photon continuum of hydrogen in the this non-radiative
shock, and that continuum from the 2s level of H is excited in a manner
similar to the n=3 levels that produce H$\alpha$.

The orientation of the dispersion axis of the COS instrument was close 
to the long direction of the filament for the longer observations (18$^\circ$ 
and 16$^\circ$ away), though it was nearly perpendicular to the filament 
(70$^\circ$) for the earlier single orbit exposure.  Because the dispersion 
direction was along the filament for 90\% of the exposure time,
we assume that the emission filled the aperture uniformly in the dispersion
direction.

The position was chosen because the fluxes were known from the single orbit spectrum
in \cite{raymond13} and because the H$\alpha$ profile of \cite{medina} provides the
proton temperature.  It also provides an electron-proton temperature ratio of $>$0.8,
consistent with the determination of \cite{ghavamian01} for a nearby position.  In addition, the proper
motions of the shocks in that region have been measured and the electron temperature
is available from X-ray spectra \citep{salvesen09, katsuda}.  IR images from Spitzer and UV
spectra from COS have been used to investigate the destruction of dust and the 
sputtering of carbon atoms from grains behind this shock \citep{sankrit, raymond13}.  Moreover,
a FUSE spectrum of a position about 2.7$^\prime$ to the NW along the same filament
provides the oxygen kinetic temperature from the O VI line profile \citep{raymond03}.

The expected line width is comparable to the instrumental profile of COS
for an extended source with the G160M grating, but no instrumental profile
was available for extended sources other than the statement by \cite{france} that the
spectral resolution is about 200 \kms.  The profile is not expected to be strictly Gaussian, 
but to be affected by the shape of the entrance aperture.  To obtain the instrument profile
for an extended source of narrow emission lines through the COS aperture, 
we observed a position in the planetary nebula NCG 6853 (The Dumbbell) 28$^\prime$$^\prime$ S 
and 2$^\prime$$^\prime$ E of the central star.  \cite{barker} reported 
emission line fluxes from an IUE spectrum at that position, and \cite{goudis} show 
[O III] line profiles of NGC6853. The two positions closest to the COS position 
have line widths (FWHM) of about 30 and 60 \kms .  An archival STIS spectrum of the central star
of NGC 6853 shows saturated absorption against the stellar continuum in the C IV lines between 
0 and -40 \kms, which probably removes the blue wings of the emission lines, so the width
is about 30 \kms, which is small compared to the COS instrumental FWHM.  At worst, the FWHM
of the emission lines could be about 40 \kms, which would mean that we overestimate the
width of the instrumental profile by 4\%. 
We find that the profiles of the C IV $\lambda$1550 doublet and the He II $\lambda$1640 line
in NGC 6853 can be fit with pairs of Gaussians.  For the C IV lines, the FWHM is 0.40 \AA\/
and the separation is 0.86 \AA , while for the He II line the best fit is FWHM = 0.47 \AA\/
and separation = 0.93 \AA .  The larger width of the He II line is at least partly due to
the fact that He II $\lambda$1640 is an unresolved multiplet with components spread over
about 0.16 \AA.

\begin{figure}
\epsscale{0.9}
\plotone{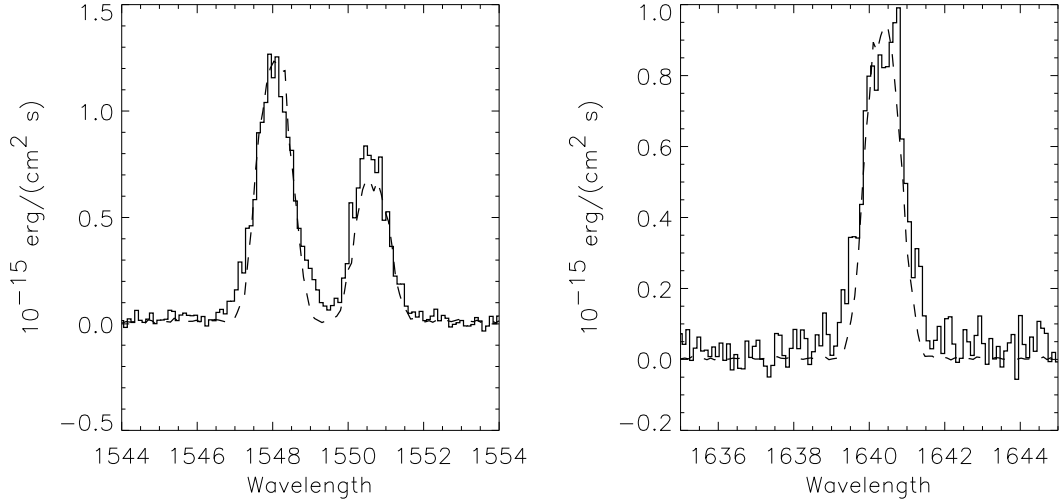}
\caption{ Observed C IV doublet and He II $\lambda$1640
lines from the Cygnus Loop (solid) compared with 
the instrument profiles derived from the spectrum of
the planetary nebula NGC 6853 (dashed).
}
\label{profcomp}
\end{figure}

\begin{figure}
\epsscale{0.9}
\plotone{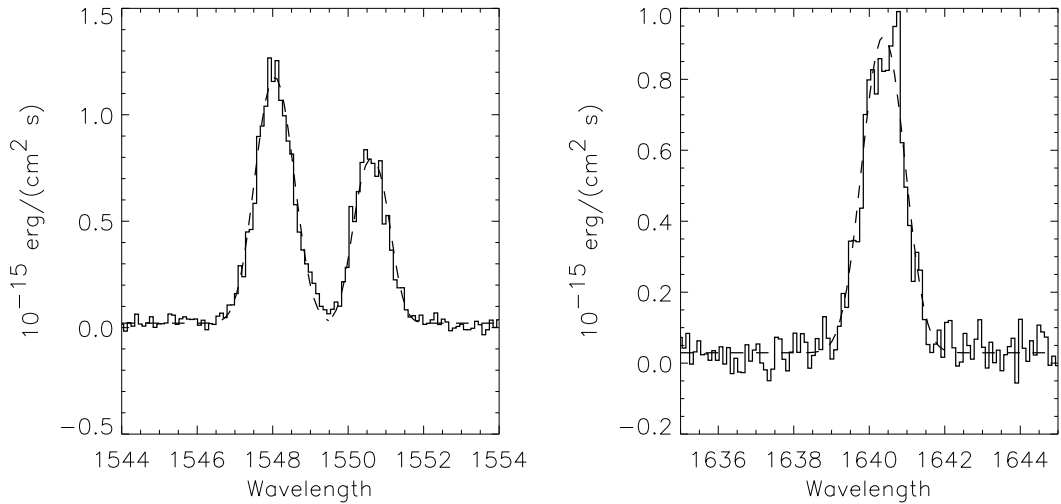}
\caption{ Observed C IV doublet and He II $\lambda$1640
observed profiles (solid) compared with best fit Gaussians 
convolved with the planetary nebula line profiles (dashed).
}
\label{fitcomp}
\end{figure}

Figure~\ref{profcomp} shows the observed line profiles along with the instrument profiles from the planetary
nebula measurements described above.  The Cygnus Loop profiles are somewhat wider than the
instrument profile, but not by much.
We fit the profiles with Gaussians convolved with the instrument profiles described above.
Figure~\ref{fitcomp} displays these fits, and Table 1 shows the results with 1-$\sigma$ errors,
along with the corresponding kinetic temperatures.  The proton and O VI temperatures from
\cite{medina} and \cite{raymond03} are included for comparison.  Note that the proton temperature
is not obtained directly from the H$\alpha$ line width.  Charge transfer between protons
and neutral atoms produces a population of neutrals given by the proton thermal distribution
convolved with the charge transfer cross section times the relative velocity \citep{cr78},
so the population of fast neutrals is similar to, but not the same as, the proton velocity distribution.
We use the model of \cite{ckr} to derive the proton temperature.

\section{Analysis}

Table 1 shows that the kinetic temperatures of helium and carbon are indistinguishable from
the proton temperature, with upper limits of 1.4 and 2.9 times the proton temperature, or
0.35 and 0.24 times the mass-proportional temperatures, respectively.
Here we discuss corrections for optical depth and Coulomb scattering used to obtain those limits.

\subsection{Optical depth effects}

The optical filaments of the Cygnus Loop and other SNRs are tangencies of the
line of sight to a thin, rippled sheet of emitting plasma \citep{hester87}.  Thus
they are essentially thin slabs of emission seen edge-on.  Resonance lines such
as the C IV doublet can have substantial optical depths along the line of sight,
but small optical depths in the perpendicular direction, so that photons are scattered out
of the line of sight \citep{long92, cornett}.  The net effect for lines of
sight near tangency is to reduce the intensity and, because the optical depth
is highest at line center, to broaden the profile \citep{raymond03}.

The ratios of the intrinsic emissivities and of the scattering cross sections for the C IV 
doublet are both 2:1, but the observed intensity
ratio is 1.65:1.  In the single
scattering limit, the intensity at any wavelength is proportional to (1-$e^{-\tau}$),
and the observed intensity ratio corresponds to optical depths at line center of about 
0.65 and 0.33 for $\lambda$1548 and $\lambda$1550, respectively. Those optical depths
imply increases of the FWHM by 12\% and 5.3\%, respectively.  We correct
the carbon kinetic temperatures for these broadenings in Table 1.  We note that the 
$\lambda$1548 line is nominally wider than the $\lambda$1550 line, though the difference 
is easily within the uncertainties.  If the apparent larger width of $\lambda$1548 were
attributed to scattering, the observed relative intensities would be close to 1:1. The optical depth 
correction for the width of the O VI line is more severe because of its larger 
optical depth, and we use the value given by \cite{raymond03}.

The He II $\lambda$1640 line is optically thin, and it is not directly affected. However, some
He II $\lambda$256 photons are absorbed and 12\% of the absorbed 
$\lambda$256 photons are converted into $\lambda$1640
photons.  Since the optical depth is largest near line center, more absorptions occur close
to line center, producing excess $\lambda$1640 photons near line center and making
the line narrower.  However, the $\lambda$256 optical depth perpendicular to 
the shock is about 6, and much larger along the line of sight.  Therefore, 
He II Ly$\beta$ photons are converted to $\lambda$1640 photons  over much 
of the line profile, and we will ignore this effect on the line width. 

\subsection{Coulomb Equilibration}

We are interested in the ion kinetic temperatures just behind the shock front,
but it takes some time for carbon to be ionized to C IV, and both He II and
C IV emit over a range of distances behind the shock determined by
the ionization times.  During those times, Coulomb collisions
will bring the kinetic temperatures toward equilibrium with the proton
temperature.  Thus we must compute the changes in the ion temperatures due
to Coulomb equilibration to see how accurately the observed kinetic temperatures
reflect the values at the shock.  Both because the ionization rates are smaller for higher ionization
states and because the Coulomb collision rate scales as $Z^2$, we expect that this
equilibration will be negligible for He II, larger for C IV and quite significant 
for O VI.

We compute the time-dependent ionization states of He, C and O using the
electron temperature from X-rays of 2$\times 10^6$ K with the ionization rates of \cite{dere07}
as compiled in version 7 of CHIANTI \citep{landi12}.  We use a proton temperature
of $1.8 \times 10^6$ K from the H$\alpha$ profile with ion-proton equilibration
rates from \cite{spitzer}.  The difference between the two assumed temperatures lies
within the uncertainties.  As a limiting case, we assume mass-proportional initial temperatures,
$T_{He} = 7.2\times 10^6$ K, $T_C = 2.2\times 10^7$ K and $T_O = 2.9\times 10^7$ K, with 
all the ions starting in the
singly ionized state.  The density does not matter, because the ionization and
Coulomb collision rates scale in the same way with $n_e$.

\begin{figure}
\epsscale{0.7}
\plotone{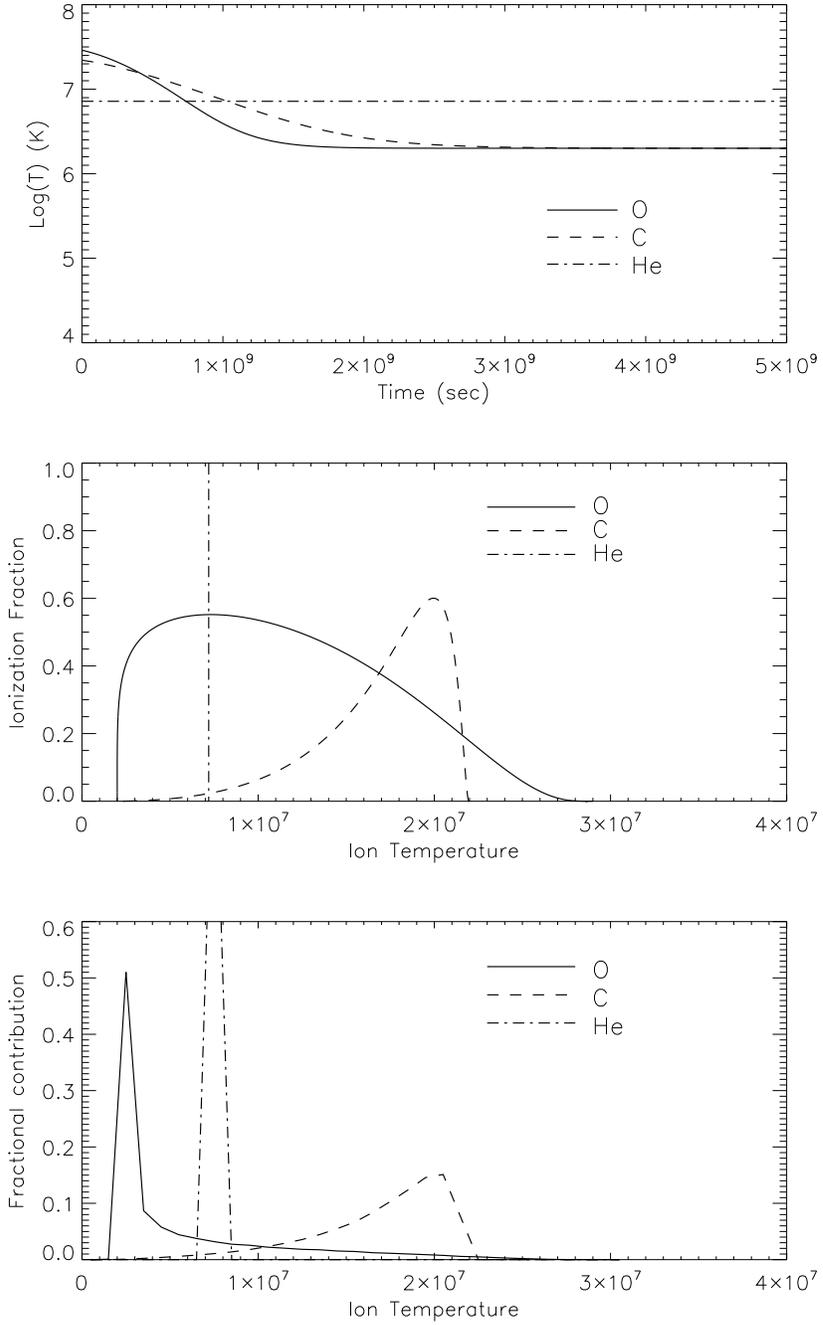}
\caption{ Top panel.  Kinetic temperatures of He II, C IV and O VI as functions of
time behind the shock front.  Middle panel.  Ionization fractions of He II, C IV and O VI as functions of
kinetic temperature.  While Coulomb collisions have no effect on helium
before He II is ionized away, and only a small effect on the the temperature
of carbon before C IV is ionized to C V, the temperature of oxygen is
driven toward the proton temperature by the time O VI is ionized.  Bottom panel.
Contribution fractions to the emission of each line in $10^6$ K bins.
}
\label{tequil}
\end{figure}

The top panel of Figure ~\ref{tequil} shows the kinetic temperatures of He II, C IV and O VI
ions as functions of time for a density of 2.0 $\rm cm^{-3}$.  It is apparent that carbon and
oxygen approach equilibrium relatively quickly because of the $\rm Z^2$ dependence of the Coulomb
collision rates.  This plot is not very useful, however,
because nearly all of the oxygen is in lower ionization states at short times and in higher ones
at long times.  The middle panel shows the ionization fractions of He II, C IV and O VI as
functions of the kinetic temperatures of those ions.  The result, as expected, is that 
the average kinetic temperature of the He II is
the same as the initial temperature.  The carbon ionization fraction peaks when the C IV kinetic
temperature had dropped noticeably, and the O VI kinetic temperature has dropped even more
sharply before the peak ionization fraction has been reached.

The bottom panel in Figure ~\ref{tequil} shows the fractional contribution to the emission for
each ion as a function of temperature in $10^6$ K bins.  The emission of He II is confined to a
single bin, while the emission of C IV is concentrated between $1.6 \times 10^7$ and $2.1 \times 10^7$ K,
with a tail at lower temperatures.  The O VI emission peaks in the bin between $2\times 10^6$ and
$3 \times 10^6$ K with a long tail at higher temperatures.  The peak of the O VI contribution in the
low temperature bin close to the proton temperature comes about because the oxygen spends a lot
of time at those kinetic temperatures, which compensates for the modest ionization fraction. 
The average temperatures are 
$7.2 \times 10^6$, $1.7 \times 10^7$ and $5.6 \times 10^6$ K for He II, C IV and O VI, respectively,
or 1.0, 0.77 and 0.2 times the mass-proportional temperatures.  We computed similar models for a
range of intial temperatures and find that the upper limits in Table 1 are compatible with intial
temperatures of 0.5, 0.3 and 0.5 times the mass proportional temperatures for He, C and O, respectively.
Smaller values would be needed to match the best fit temperatures. 

The fact that C IV and O VI are formed over a
range of temperature means that their line profiles are the sum of Maxwellians, 
so they will deviate from Gaussians.  The C IV line is formed over a range of about 25\% in temperature
or 13\% in thermal speed, so detecting the departure from Maxwellian would require
better data than the spectrum presented here.  The O VI profile from a shock with mass-proportional
heating would give a core with a temperature just above the proton temperature plus broad
wings containing a large part of the flux.  While the some of the profiles shown in 
\cite{raymond03} show a hint of a wing on the red side, there is no indication of the
strong wings predicted by a model with mass-proportional temperatures.  

Thus we can ignore Coulomb collisions for the He II line, and the C IV lines widths are
affected at the 12\% level, but the O VI profiles presented by
\cite{raymond03} are severely affected.  In the faster shocks in DEM L71 \citep{ghavamian07b}
and SN1006 \citep{raymond95, korreck04}, the higher electron and ion temperatures
greatly reduce the ionization time scales and increase the Coulomb collision time
scale, so that Coulomb collisions can be neglected even for O VI.

\subsection{Interpretation} 

First we discuss the kinetic temperatures that might be expected in a 
collisionless shock.  To first order, such a shock simply thermalizes
3/4 of the shock speed independently for each species, leading to 
mass-proportional ion temperatures, approximately as seen in the 3000 \kms shock
in SN1006 \citep{raymond95, korreck04}. 

The situation is more complicated for species that enter the shock as neutral atoms
or in dust grains.  Much of the helium is neutral, and about 75\% of the carbon is
in the form of PAHs or dust \citep{raymond13}.  The PAHs dissociate very quickly behind
the shock and contribute to the observed line profile \citep{micelotta}, while dust grains
sputter away more gradually over tens of arcseconds at the distance of the Cygnus Loop 
\citep{raymond13} and do not contribute 
much to the profile at the shock.  Neutrals that pass through the shock and become 
ionized downstream are expected to behave as pickup ions \citep{raymond08}.  For 
quasi-perpendicular shocks, the velocity component along the field is conserved while
the component perpendicular to the field forms a ring beam in velocity
space.  The unstable ring distribution quickly scatters into a hollow bispherical
shell in velocity space by emitting and absorbing Alfv\'{e}n waves 
\citep{williamszank, isenberg}.  The distribution then scatters
into a Maxwellian, though the temperatures parallel and perpendicular to the field
may differ.  If the distribution is too anisotropic, either firehose or mirror instabilities
will drive the anisotropy toward marginal instability \citep{maruca}.  Such a
pickup ion process in the partially neutral hydrogen
provides one of several possible interpretations for the non-Maxwellian H$\alpha$
profile observed at knot ''g" in Tycho's supernova remnant \citep{raymond10}, and it is
responsible for the motion of oxygen ions along field lines seen in images of the sungrazing
comet C/2011 W3 (Lovejoy) \citep{raymond14}.  The poor fit of the He II profile to a
Gaussian velocity distribution, $\chi^2 = 2.4$, might suggest a pickup ion
contribution added to a Maxwellian, but the data do not warrant a strong conclusion.
We note that the sharp peak to the right of the centroid is much narrower than
the instrumental resolution and must therefore be a result of the modest number of
counts in the line.

The comet observations also confirm that
a significant fraction of the energy in the ring distribution can be lost to waves,
depending on the Alfv\'{e}n speed and angle between the flow and the magnetic field
\citep{williamszank}.  For the shock observed here, the post-shock Alfv\'{e}n speed 
is around 20-30 \kms , and the loss to waves should be small \citep{raymond08}.
The final temperature of the pickup ions depends on the angle 
between the field and the shock, and for quasi-perpendicular shocks it will be between 0.5 and 1
times the mass-proportional temperature.

Many of the C and O atoms are already ionized when they pass through the shock.  Because of their high
mass to charge ratios, these ions are not slowed as much by the electrostatic field in the shock
wave as are the protons.  According to \cite{fuselier} they will be about 25\% hotter than the protons.
\cite{leewu} and \cite{zimbardo} also present models for strong preferential heating of heavy ions
in shocks.  Those models predict very high ratios of perpendicular to parallel temperatures,
$T_\bot / T_\|$.  Depending upon the angle between the magnetic field and our line of sight,
we could in principle observe mainly $T_\|$ and therefore a low temperature.  However, large
$T_\bot / T_\|$ would generate mirror mode instabilities and, for the high plasma $\beta$ in this
shock, sharply reduce the anisotropy \citep{maruca}.
Solar wind observations of modest Mach number shocks show some preferential heating of
heavy ions, but there is little correlation with shock parameters \citep{korreck07}.

There is also transfer of energy among species by plasma waves and turbulence at the 
shock.  Waves can be excited upstream of the shock jump by streaming instabilities generated
either by particles reflected by the shock or by cosmic rays \citep{thomsen90, laming14}.  For example,
\cite{ghavamian07} and \cite{rakowski08} show that lower hybrid waves in a cosmic ray precursor 
can heat the electrons to $T_e \simeq T_p$ in shocks at about the speed of the one observed here.    

Our observations give upper limits to He and C temperatures of 1.4 and 2.9 times the proton temperature
found by \cite{medina}.  After considering Coulomb collisions, the upper limits on the initial temperatures
are 2 and 4 times the proton temperature, and the O VI profile of \cite{raymond03} gives an upper limit of 8 times
the proton temperature.  This contrasts with the nearly
mass proportional values (4, 12 and 16) seen in the 2000-3000 \kms shock in SN1006 \citep{raymond95, korreck04},
and it is similar to the indications of nearly complete electron-ion equilibration in the Cygnus
Loop compared with $T_e/T_i \simeq 0.05$ in SN1006 \citep{ghavamian02, vanadelsberg}.  It
indicates rapid temperature equilibration by plasma waves and turbulence in the Mach 35 Cygnus
Loop shock, while the equilibration in the Mach 200-300 SN1006 shock is ineffective.

The effective equilibration of the particle species has an important implication for the derivation
of shock speeds from H$\alpha$ line widths. As shown by \cite{ckr} and \cite{vanadelsberg}, if
the energy dissipated in the shock is shared among electrons and ions, a higher shock speed is needed to
account for a given line width.  On the other hand, if helium is efficiently equilibrated, the
proton temperature behind a shock of a given speed is increased.  \cite{medina} assumed that 
helium remained mostly neutral through the region where the H$\alpha$ line is formed and that
the $T_e$ and $T_i$ are equal, so that the post-shock temperature is given by 1/2 the thermalized 
proton speed.  On the other hand, we find that helium is brought to close to the ion temperature
close to the shock, in which case the kinetic energy of the helium is shared with the protons.
While \cite{medina} found a shock speed of 405 \kms, a shock speed as low as $\sim 360$ \kms could
account for the H$\alpha$ profile if $T_H = T_{He}$ where the H$\alpha$ line forms.  (We assume
a helium abundance of 10\% by number.  If most of the helium is neutral and it is mostly ionized
only after the hydrogen is ionized, the effect will be smaller.) When that
lower shock speed is combined with the proper motion measured by \cite{salvesen09}, the distance
to the Cygnus Loop is reduced from 890 pc to 800 pc.  That partially alleviates the disagreement
between the proper motion distance and the upper limit of 640 pc to the distance from the detection
of SNR absorption lines \citep{blair09} in the spectrum of a background star, but it still leaves 
a substantial discrepancy.  Of course, if a significant fraction of the shock energy goes into 
cosmic rays, a correspondingly higher shock speed would be required.  \cite{shimoda} point out that
comparison of the shock speed derived from the proper motion with the shock parameters determined from
the shock jump conditions can be misleading, as a shock in a realistic ISM density distribution
is somewhat oblique at most positions.  However, for the shock observed here, \cite{medina} measured the 
offset between the velocity centroids of the broad and narrow H$\alpha$ to be only 9 \kms , so 
the obliquity is very small.  

\section{Summary}

We have measured He and C kinetic temperatures behind a moderate velocity shock in the Cygnus Loop.
Unlike the fast shock in SN1006, where the temperatures are mass-proportional, in the Cygnus Loop
they are nearly the same.  We have shown that Coulomb equilibration is not important for the He II
line, and it affects the C IV lines at the 33\% level,
so the equilibration must occur at the shock front by means of plasma turbulence.  However,
Coulomb collisions severely affect the kinetic temperature of O VI if it is initially higher
than the proton temperature, so that the C IV and He II lines provide better constraints on temperatures
at the shock than do the O VI lines for this relatively slow SNR shock. The rapid equilibration of
helium suggests that the post-shock protons are partly heated by helium, which would reduce the
shock speed required, and hence the distance inferred from proper motions, by as much as 11\%.
It is important that the the thermal equilibration of protons, electrons and helium be considered
when converting H$\alpha$ line widths of Balmer line filaments into shock speeds.

The degree of ion-ion equilibration and the related heating processes may be important for determining the
efficiency of injection of the different species into the cosmic ray acceleration process.  Besides
that, it is potentially useful as a diagnostic for the plasma waves that produce the shock transition
in a collisionless shock.

We determined the instrumental profile of the COS spectrograph for an extended source with the G160M 
grating by observing a planetary nebula, and this may be useful for other studies of extended sources
that fill the COS aperture.  The $\simeq$ 200 \kms\/ wide instrumental profile can be described as 
a pair of Gaussians with FWHM = 0.40 \AA\/ separated by 0.86 \AA .

\bigskip
The authors thank the referee for requesting an additional plot, which revealed
a numerical error in the average temperatures, and for other useful suggestions.
This work was performed under grants HST-GO-12885 and HST-GO-13436 to the 
Smithsonian Astrophysical Observatory. RJE was supported by NASA contract
NAS8-03060.   JR and PG thank Lorentz Center Workshop on Particle Acceleration from 
the Solar System to Cosmology for useful discussions.
  
{\it Facilities:} \facility{HST (COS)}



\begin{table}
\begin{center}

\centerline{Table 1}
\vspace{2mm}
\centerline{Fit Parameters}
\vspace{2mm}
\centerline{Temperatures in units of $10^6$ K}

\vspace{5mm}
\begin{tabular}{ l c c c c }
\hline \hline
Line                 &   FWHM (\kms)  & $T_{kin}$ ($10^6$ K)  & $T_{corr}$ ($10^6$ K) & $\chi^2$/DOF   \\
\hline
                     &                &                        &                   &        \\
H I   $\lambda$6563  & 254$_{-16}^{+16}$$^a$ &  1.8$_{-0.3}^{+0.3}$$^b$ &                   &       \\
He II $\lambda$1640  & 106$_{-40}^{+39}$ &  2.0$_{-1.2}^{+1.8}$ & 2.0$_{-1.2}^{+1.8}$ &  2.46  \\
C IV  $\lambda$1548  & 106$_{-42}^{+25}$ &  5.9$_{-3.7}^{+3.2}$ & 4.7$_{-3.0}^{+2.5}$ & 1.33 \\
C IV  $\lambda$1550  & 72$_{-71}^{+35}$ &  2.7$_{-2.7}^{+3.2}$ & 2.4$_{-2.4}^{+2.8}$ & 1.43 \\
O VI  $\lambda$1032  & 120             &  5.1$^c$           & $<$3.2$^c$           &    \\
\hline

\end{tabular}
\end{center}
a  H$\alpha$ line width from \cite{medina}  \\
b  Hydrogen temperature based on model proton temperature leading to observed line width after 
charge transfer \citep{medina} \\
c  From \cite{raymond03}. The temperature is $3.2 \times 10^6$ K after correction for optical depth effects.
\end{table}

\newpage

\end{document}